\documentclass{aastex63}

\hypersetup{linkcolor=red,citecolor=green,filecolor=cyan,urlcolor=magenta}
\usepackage{mathrsfs}  

\submitjournal{ApJ}

\shorttitle{BSSs in NGC6256}
\shortauthors{Cadelano et al.}
\graphicspath{{./}{figures/}}

\begin{document}

\title{Discovery of a double sequence of blue straggler stars in the core-collapsed globular cluster NGC 6256}

\correspondingauthor{Mario Cadelano}
\email{mario.cadelano@unibo.it}
\author[0000-0002-5038-3914]{Mario Cadelano}
\affil{Dipartimento di Fisica e Astronomia, Università di Bologna, Via Gobetti 93/2 I-40129 Bologna, Italy}
\affil{INAF-Osservatorio di Astrofisica e Scienze dello Spazio di Bologna, Via Gobetti 93/3 I-40129 Bologna, Italy}

\author[0000-0002-2165-8528]{Francesco R. Ferraro}
\affil{Dipartimento di Fisica e Astronomia, Università di Bologna, Via Gobetti 93/2 I-40129 Bologna, Italy}
\affil{INAF-Osservatorio di Astrofisica e Scienze dello Spazio di Bologna, Via Gobetti 93/3 I-40129 Bologna, Italy}

\author[0000-0003-4237-4601]{Emanuele Dalessandro}
\affil{INAF-Osservatorio di Astrofisica e Scienze dello Spazio di Bologna, Via Gobetti 93/3 I-40129 Bologna, Italy}

\author[0000-0001-5613-4938]{Barbara Lanzoni}
\affil{Dipartimento di Fisica e Astronomia, Università di Bologna, Via Gobetti 93/2 I-40129 Bologna, Italy}
\affil{INAF-Osservatorio di Astrofisica e Scienze dello Spazio di Bologna, Via Gobetti 93/3 I-40129 Bologna, Italy}

\author[0000-0002-7104-2107]{Cristina Pallanca}\affil{Dipartimento di Fisica e Astronomia, Università di Bologna, Via Gobetti 93/2 I-40129 Bologna, Italy}
\affil{INAF-Osservatorio di Astrofisica e Scienze dello Spazio di Bologna, Via Gobetti 93/3 I-40129 Bologna, Italy}

\author[0000-0003-4746-6003]{Sara Saracino}
\affil{Astrophysics Research Institute, Liverpool John Moores University, 146 Brownlow Hill, Liverpool L3 5RF, UK}
\affil{INAF-Osservatorio di Astrofisica e Scienze dello Spazio di Bologna, Via Gobetti 93/3 I-40129 Bologna, Italy}

\begin{abstract}
We used a combination of high-resolution optical images acquired with the Hubble Space Telescope and near-IR wide-field data to investigate the stellar density profile and the population of blue straggler star (BSS) in the Galactic globular cluster NGC6256, with the aim of probing its current stage of internal dynamical evolution. We found that the inner stellar density profile significantly deviates from a King model while is well reproduced by a steep cusp with a power-law slope $\alpha_{\rm cusp}=-0.89$, thus implying that the cluster is currently in the post core-collapse (PCC) phase. This is also confirmed by the very high segregation level of the BSS population measured through the $A^+_{rh}$ parameter. We also found that the distribution of BSSs in the color-magnitude diagram is characterized by a collimated blue sequence and a red more sparse component, as already observed in other three PCC clusters. The comparison with appropriate collisional models demonstrates that the vast majority of the BSSs lying along the collimated blue sequence is consistent with a generation of coeval (1 Gyr-old) stars with different masses originated by an event that highly enhanced the collisional rate of the system (i.e. the core collapse). This study confirms that the segregation level of BSSs is a powerful dynamical diagnostic also of star cluster in a very advanced stage of dynamical evolution. Moreover, it pushes forward the possibility of using the morphology of the BSS in the color-magnitude diagram as a tracer of the core-collapse and subsequent dynamical evolutionary phases.
\end{abstract}



\section{Introduction}
\label{sec:intro}
Blue straggler stars (BSSs) are an exotic population of stars located
in a more luminous and bluer region than main-sequence (MS) turn-off
(TO) in the color-magnitude diagram (CMD) of stellar systems
\citep[e.g.,][]{sandage53, ferraro97, ferraro03, ferraro12, ferraro18,
  piotto04, leigh07, boffin15}. They are core hydrogen-burning stars
more massive \citep[e.g.,][]{fiorentino14,raso19} than MS-TO objects
and {thought to be the outcome of two main processes:
  mass-transfer and/or coalescence in binary systems (hereafter,
  MT-BSSs; \citealp{mccrea64}), and (2) direct stellar collisions
  (COL-BSSs; \citealp{hills76}). Other processes, such as merger of
  close binary systems, possibly induced by Kozai-Lidov mechanisms in
  hierarchical triple systems, can also play a role in generating BSSs
  \citep{perets09}. While the formation channels requiring binary
  systems are common in all stellar environments and dominant in
  low-density conditions, such as the peripheries of globular clusters
  (GCs), open clusters, dwarf galaxies and the Galactic field, the
  collisional channel requires very high-density environments.  Hence,
  the inner cores of GCs offer the most favorable conditions for
  direct stellar collisions to occur, although the MT channel is
  considered the most prolific one \citep[e.g.][]{davies04, knigge09,
    leigh13}.}  Since BSSs are ubiquitous found in all GCs and are
more massive than the average \citep[e.g.][]{ferraro18,rain21}, they
are used as powerful tools to study star cluster internal dynamics
\citep[e.g.][]{ferraro12, ferraro18, ferraro19, ferraro20,
  alessandrini16, lanzoni16, dresbach22}. In particular,
\citet{ferraro18} studied about one third of the entire GC population
in our Galaxy and found a tight correlation between the number of
relaxation times suffered by each cluster since formation ($N_{\rm
  relax}$) and the value of the $A^+_{rh}$ {parameter.  The latter
  is defined as the area between the cumulative radial distribution of
  BSSs and that of a reference population (such as MS, giant and
  horizontal branch stars), within a half-mass radius from the center
  of the system \citep{alessandrini16}.} Its value thus measures the
level of BSS central segregation with respect to ``standard''
(lighter) stars, which progressively increases while the host cluster
dynamically evolves. Large values of $A^+_{rh}$ are therefore measured
for GCs in late stages of their dynamical evolution, while small
values (down to zero) are found in dynamically young systems, where
dynamical friction has not been effective yet in segregating BSSs
toward the center.

The distinction between BSSs formed through MT and collision would
provide useful information about the collision rate of the cluster,
the fraction of binaries and their impact on internal dynamics, and
also insights on the formation and evolution of these exotic objects.
Unfortunately, BSSs formed through different channels are hardly
distinguishable from their photometric properties. The only exception
known so far could be the presence of two distinct BSS sequences in
the CMD of post core-collapse (PCC) clusters\footnote{ {The core-collapse is a characteristic phase of the internal dynamical
  evolution of collisional stellar systems.  It is the result of the
  continuous kinetic energy transfer from the inner regions to the
  outskirts that leads to a runaway contraction of the core, with a
  substantial increase of its density. PCC clusters are commonly
  recognized from the presence of a steep power-law cups in the
  innermost portion of the density profile \citep{meylan97}.}}.  This
feature has been first discovered in the GC M30 \citep{ferraro09}, and
it has then been identified in a few additional systems: NGC 362
\citep{dalessandro13}, NGC 1261 \citep[][but see
  \citealp{raso20}]{simunovic14}, and M15 \citep{beccari19}\footnote{A
detection of a double BSS sequence in the Large Magellanic Cloud GC
NGC 2173 has been claimed by \citep{li18a}, but this is most likely an
artifact of field star contamination \citep{dalessandro19a,
  dalessandro19b}}.  All these clusters show in the CMD a narrow
sequence of blue BSSs, separated through a clear-cut gap from a more
scattered red BSS population. The red sparse sequence is not
reproducible with collisional isochrones \citep{sills09} of any age,
while it is in good agreement with the CMD location of MT-BSS models
\citep{xin15}, thus suggesting that it is likely populated by BSSs
formed through the MT channel. On the other hand, the blue sequence is
inconsistent with the MT-BSS models of \citet{xin15} and is well
reproduced by collisional isochrones, thus suggesting that it is
likely populated by COL-BSSs. Interestingly, Monte Carlo simulations
by \citet{jiang17} show that the blue sequence could be also
contaminated by MT-BSSs. However, its narrowness and the clear-cut
separation from the red sequence appear to be inconsistent with a
formation from MT activity in binary systems, which extends over long
time scales. On the other hand, the results obtained by
\citet{jiang17} could explain the presence of the few W-Uma stars
(contact binaries)
detected along the blue sequence of M30 and NGC 362 \citep{ferraro09,
  dalessandro13}. The narrowness of the blue sequence
suggests that it is more likely composed of a population of coeval BSSs with
different masses, all formed over a relatively short period of time.
Based on this evidence, \citet{ferraro09} suggested that the origin of
the observed blue sequence is related to the core-collapse (CC) event which enhanced the cluster collision rate in inner regions over a short time scale. This working hyphothesis was later confirmed by
numerical simulations specifically performed to reproduce the bimodal
BSS distribution of M30 (see \citealp{portegies19}). In addition, as a
matter of fact, the three clusters where the double BSS sequence has
been firmly identified so far (i.e. M30, NGC 362 and M15) already
experienced the CC phase (see \citealp{harris96}, 2010 edition).
In light of all this, \citet{ferraro09} suggested that the properties
of the blue narrow sequence could be also used to date back the CC
epoch: as the MS-TO luminosity is a proxy of the cluster age, so the
luminosity of the blue BSS sequence (or its extension in luminosity)
can provide an estimate of the epoch when the system experienced a
significant enhancement of its collision rate, which triggered the
formation of COL-BSSs that we now observe aligned in the CMD.

In this paper we present the discovery of a double sequence of BSSs in
the GC NGC 6256. This is a dense cluster ($\log{\rho_0}\approx 5.9$ in
units of $M_\odot$ pc$^{-3}$; \citealp{baumgardt18}) located in the
Galactic bulge, at a distance of 6.8 kpc from the Sun
\citep{cadelano20} and characterized by a relatively low metal content
compared to typical bulge GCs ([Fe/H]$=-1.6$; \citealp{vasquez18}). In
\citet{cadelano20}, we applied a well-tested procedure
\citep[see][]{pallanca19,pallanca21} to quantify the differential
reddening affecting the cluster and we derived a high-resolution
exctiction map (see Figure~3 in \citealt{cadelano20}): color excess
variations as large as $\delta E(B-V)\sim0.51$ mag were measured in
the relatively small $160\arcsec \times 160\arcsec$ field of view
(FOV) sampled by the adopted data-set. The differential reddening
corrected CMD allowed the authors to derive an age of $13\pm0.5$ Gyr.
The analysis of the CMD also revealed a severe contamination by field
interlopers along the cluster's main evolutionary sequences.

In the comprehensive analysis of GC surface brightness profiles by
\citet{trager93, trager95}, NGC 6256 is classified as a PCC cluster,
due to the presence of a central surface brightness cusp. However, the
cluster structural properties based on surface brightness profiles
suffer from the so-called ``shot-noise bias'', due to the stochastic
and sparse presence of luminous stars, which can significantly
displace the surface brightness peak from the true location of the
cluster gravitational center and alter the shape of the surface
brightness profile with respect to the true density distribution
\citep[see, e.g.][]{noyola06, cadelano17, lanzoni19}.

Recently, \citet{cohen21} determined the cluster's structural
parameters from the star count density profile. They derived a cluster
gravitational center by fitting ellipses to the isodensity contours,
and found a value in agreement, within the uncertanties, with the one
quoted in \citet{cadelano20}, which was calculated through iterative
re-centering of the stellar average coordinates. They also determined
the cluster's structural parameters through \citet{king66} model
fitting of the density profiles obtained by combining the same
high-resolution observations used throughout this work (see
Section~\ref{sec:obs}) to sample the central regions, with Gaia DR2
photometry \citep{gaia18} to sample the cluster's outskirts. The
authors discuss that the best-fit model is unable to properly
reproduce the observed stellar density in the cluster's innermost
regions, which is expected in the case of a very dynamically evolved
and core-collapsed cluster.

This paper is structured as follows: in Section~\ref{sec:obs} we
introduce the adopted data-set and the data reduction procedures; in
Section~\ref{sec:profile} we determine the cluster's stellar density
profile and investigate its dynamical status; {in
  Section~\ref{sec:bss} we discuss the proper motion (PM) analysis and
  present the discovery of a double sequence of BSSs.} Finally, we
summarize the result and draw our conclusions in
Section~\ref{sec:conclusions}.

\section{Observations and data analysis}
\label{sec:obs}
{\it High-resolution data-set:} this work is mainly based on two
data-set of optical images obtained with the {\it Hubble Space
  Telescope} (HST). The first data-set was obtained using the UVIS
channel of the Wide Field Camera 3 (WFC3) under GO 11628 (PI:
Noyola). It consists of 3 images acquired with the F555W filter and
exposure time of 360 s, and 3 images in the F814W filter with exposure
time of 100 s. The adopted procedures of data reduction, astrometry
and calibration are described in detail in \citet{cadelano20}, where
it is also discussed the determination of a high-resolution
differential reddening map in the direction of the system. The second
data-set was obtained under GO 15065 (PI: Cohen) using the Wide Field
Camera (WFC) of the Advanced Camera for Surveys (ACS). It consists of
4 images acquired with the F606W filter with exposure time of 498 s
and 4 images acquired with the F814W filter with exposure time of 509
s. The data reduction was performed using a standard approach with
DAOPHOT and ALLFRAME packages \citep{stetson87,stetson94}, in a
similar way as described in \citet{cadelano20} in the case of the WFC3
data-set. The FOVs of the two HST data-sets overlap and are separated
by a temporal baseline of 3605 days ($\sim10$ yr), thus allowing the
measure of stellar PMs (see Section~\ref{sec:pm}). All the HST
magnitudes used hereafter throughout the work are corrected for the
effects of differential reddening.

{\it Wide-field data-set:} to dermine the stellar density profile
along the entire radial extension of the cluster, we complemented the
high-resolution HST data, with a set of wide-field near-infrared
images obtained as a part of the VVVX survey
\citep{minniti_vvvx}. This data-set was acquired with the VISTA
InfraRed CAMera (VIRCAM) mounted on the VISTA-ESO telescope. It is
composed of 2 images obtained with the J filter and exposure time of
120 s, 1 image with the H filter and exposure time of 48 s and,
finally, 1 image with the Ks filter and exposure time of 16 s. Also in
this case, the data reduction was performed using a standard approach
suitable for ground-based observations with DAOPHOT and ALLFRAME
packages \citep[see e.g.][]{cadelano20IMF}. The resulting catalog was
astrometrized using the stars in common with {the Gaia Data
  Release 3 (DR3) catalog (Gaia Collaboration, 2022, in preparation)}
and the instrumental magnitudes were calibrated using the stars in
common with the catalog by \citet{valenti07}, obtained with the same
filters and in an overlapping region of the sky.

\begin{figure}[h] 
\centering
\includegraphics[scale=0.4]{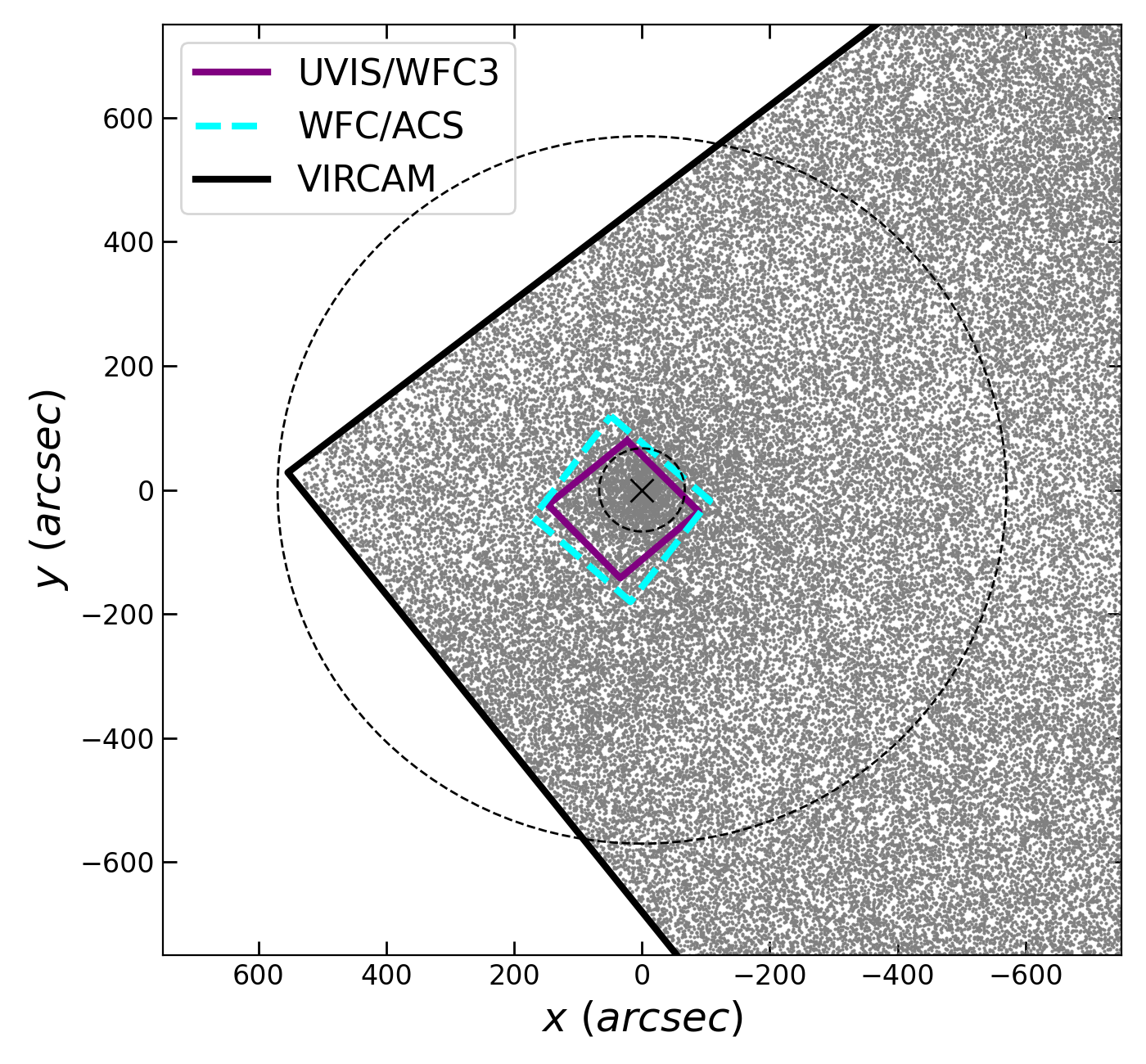}
\caption{FOVs covered by the data-sets used in this work, plotted with
  respect to the cluster's center {(black cross)} quoted in
  \citet{cadelano20}. The gray dots are the observed stars. {The
    violet and cyan thick lines mark} the FOVs of the WFC3 and ACS
  data-sets, respectively. The black solid line marks the edge of a
  portion of the wide-field VIRCAM data-set. The inner and outer
  dashed circles are the cluster's half-mass and tidal radii,
  respectively, as determined in Section \ref{sec:profile}.}
\label{fig:map}
\end{figure}

\begin{figure}[h] 
\centering
\includegraphics[scale=0.5]{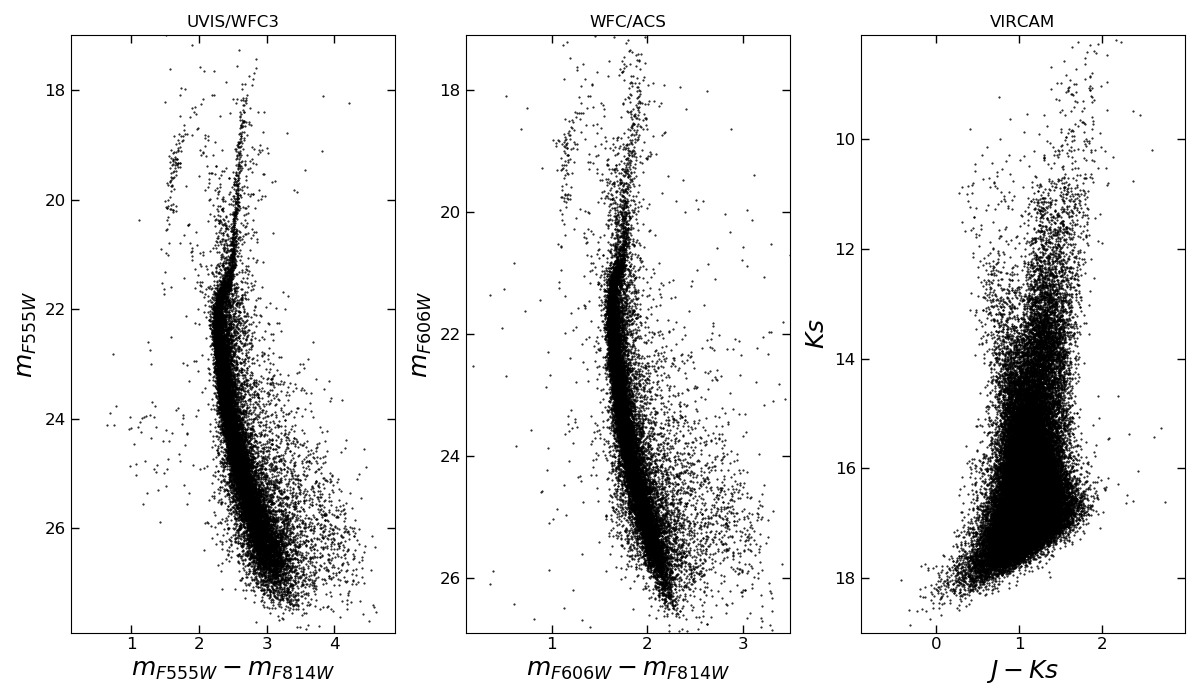}
\caption{{\it Left panel:} ($m_{\rm F555W}, m_{\rm F555W}-m_{\rm
    F814W}$) CMD of NGC 6256 as obtained from the first epoch, WFC3
  observations \citep[from][]{cadelano20}. {\it Middle panel:}
  ($m_{\rm F606W}, m_{\rm F606W}-m_{\rm F814W}$) CMD obtained from the
  second epoch, ACS observations. Both first and second epoch magnitudes are corrected for differential reddening. {\it Right panel:} ($J,J-Ks$) CMD of
  the cluster obtained from the VIRCAM observations.}
\label{fig:3cmd}
\end{figure}

The map of the FOVs covered by all the data-sets used in this work is
reported in Figure~\ref{fig:map}, while the corresponding CMDs are
shown in Figure~\ref{fig:3cmd}.

\section{Stellar density profile and structural parameters}
\label{sec:profile}
PCC clusters are typically characterized by the presence of a central
density cusp usually shaped like a power-law, rather than the flat
density core behavior predicted by \citet{king66} models and typically
observed in less dynamically evolved GCs
\citep[e.g.,][]{ferraro03_6752,  lanzoni07, vesperini10, dalessandro13, miocchi13, zocchi16}.  Based on the shape of its surface density distribution,
NGC 6256 has been classified as a PCC system \citep{trager93,
  trager95}.  In this section, we determine the cluster's density
profile from star counts to accurately estimate its structural
properties and reliably confirm its core-collapsed nature.

We adopted the gravitational center derived in \citet{cadelano20}: $\alpha = 16^{\rm h} 59^{\rm m}
32.668^{\rm s}$ and $\delta = -37^{\circ} 07\arcmin 15.139\arcsec $, 
  with an uncertainty of $\sim 0.4\arcsec$. As discussed in many previous papers 
\citep[see][]{ferraro99,ferraro03_6752,ibata09,lanzoni07}  
the accurate determination of the center of gravity is  crucial in order to 
properly characterize the star density profile, especially in high density clusters.
The projected stellar density profile has been determined following the
procedure fully described in \citet[][see also, e.g.,
  \citealp{cadelano17, lanzoni19, raso20}]{miocchi13}. To sample the
inner $\sim100\arcsec$, we used the WFC3 data-set.
For the cluster's outer regions, we used the wide-field VIRCAM
data-set. The choice of using a near-IR data-set, instead of the
publicly available Gaia DR3 catalog, is motivated by the fact that NGC
6256 is located in a region of the sky severely affected by
differential reddening \citep[see][]{cadelano20}. {Hence, the
  adoption of a catalog of stars based on optical photometry, such as
  the Gaia one, could introduce significant alterations of the density
  profile with respect to its true shape, and thus lead to a wrong
  determination of the cluster structural parameters (see the case of
  M71 discussed in \citealt{cadelano17}).}

\begin{figure}[h] 
\centering
\includegraphics[scale=0.3]{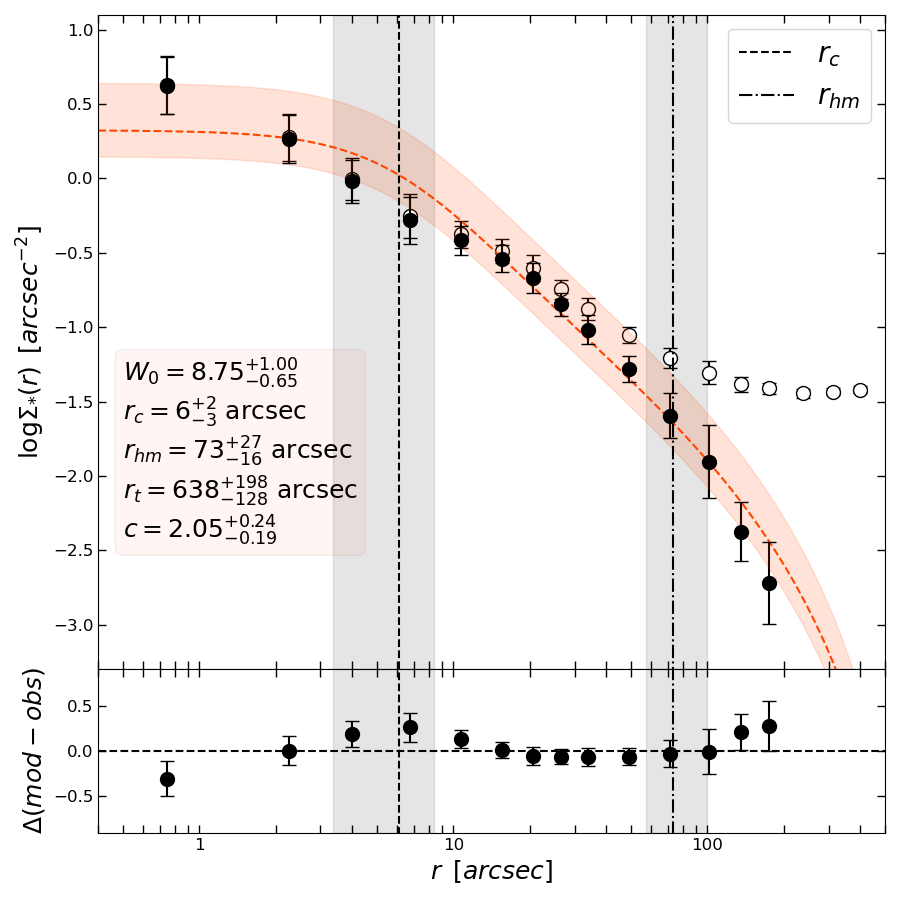}
\includegraphics[scale=0.3]{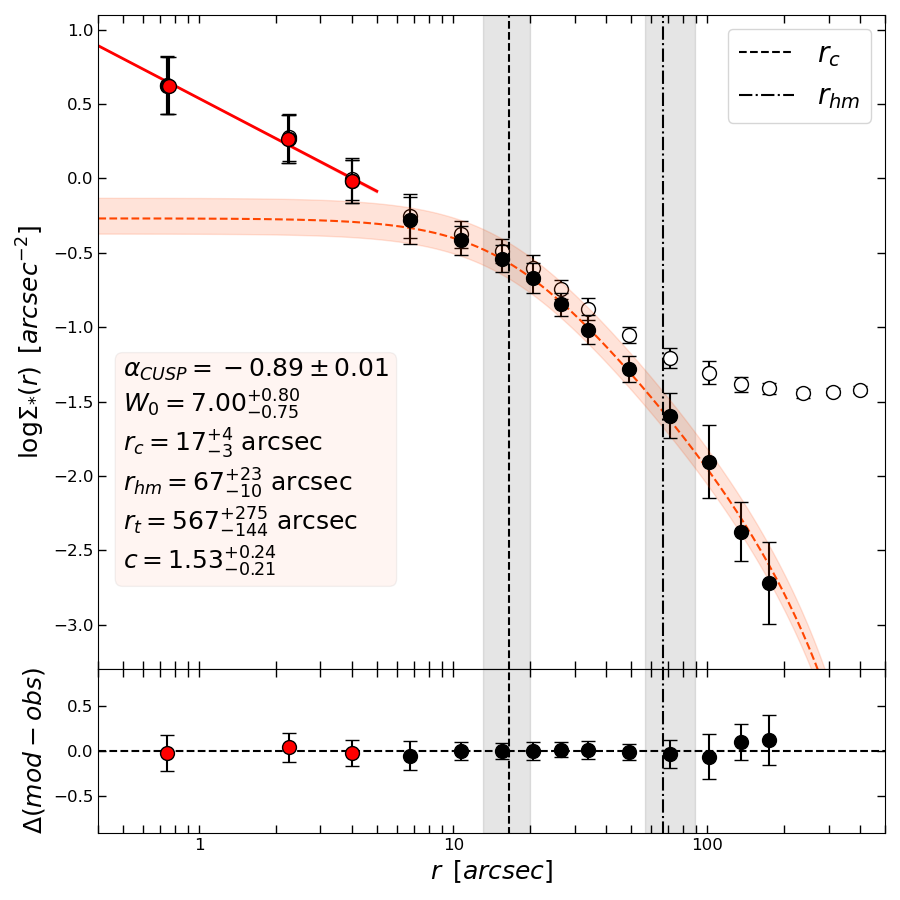}
\caption{{\it Left-hand panel:} Observed (empty circles) and
  background-subtracted (filled circles) density profile of NGC 6256.
  The dashed red curve is the best-fit King model to the cluster
  density profile and the red stripe marks the envelope of the
  $\pm1\sigma$ solutions. The dashed and dot-dashed vertical lines
  mark the best-fit cluster's core and half-mass radii, respectively,
  and their corresponding $1\sigma$ uncertainties are represented with
  the gray stripes.  The best-fit values of some structural parameters
  (see text) are also labeled. The bottom panel shows the residuals
  between the best-fit King model and the cluster density
  profile. {\it Right-hand panel:} Same as in the left panel, but with
  the three innermost points colored in red to highlight the presence
  of a stellar density cusp. These points have been fitted with the
  power-law function shown as a red solid line, having slope
  $\alpha_{\rm CUSP}$ labelled in the figure legend. The best-fit King
  model to the profile obtained by excluding the three inner points
  (black solid circles) is shown with the red dashed curve and red
  stripe, and its corresponding parameters are labelled in the
  legend.}
\label{fig:densprof}
\end{figure}

In the case of the region sampled by the HST data-set, we divided the
FOV into 12 concentric annuli out to $120\arcsec$ from the cluster's
center, each one divided into four sub-sectors. For each sub-sector,
we then counted the number of stars having magnitude $15.5<m_{\rm
  F814W}<19.8$ (to avoid possible biases due to saturation and
incompleteness of the brightest and faintest stars, respectively) and
divided it by the sampled area. The resulting density in each annulus
is the mean of the values measured in each sub-sector and the
uncertainty is the standard deviation. We repeated the same procedure
for the VIRCAM wide-field data-set, dividing the FOV in 8 concentric
annuli from $38\arcsec$ out to $450\arcsec$ from the cluster's center,
and considering only stars having magnitude $11.5<K<16.5$.  The three
innermost radial bins ($38\arcsec<r<120\arcsec$) are also sampled with
HST catalog and thus they have been used to vertically rescale the
external (VIRCAM) profile to match the inner (HST) one.  The resulting
density profile is shown in Figure \ref{fig:densprof} (empty circles).
The apparent external flattening is due to the Galactic field
contribution, which is negligible with respect to the cluster density
in the inner regions, but becomes dominant at large distances from the
center.  {The average of the three most external points in linear
  (instead of logarithmic) units has been adopted as Galactic field
  density, and then subtracted from the observed distribution to
  obtain the background-decontaminated density profile of NGC 6256
  (filled circles in Figure~\ref{fig:densprof}).}  The density profile
clearly shows a progressive increase, instead of a flat behavior,
toward the cluster center, which is the typical feature expected for a
core-collapsed GC.

\begin{figure}[h] 
\centering
\includegraphics[scale=0.2]{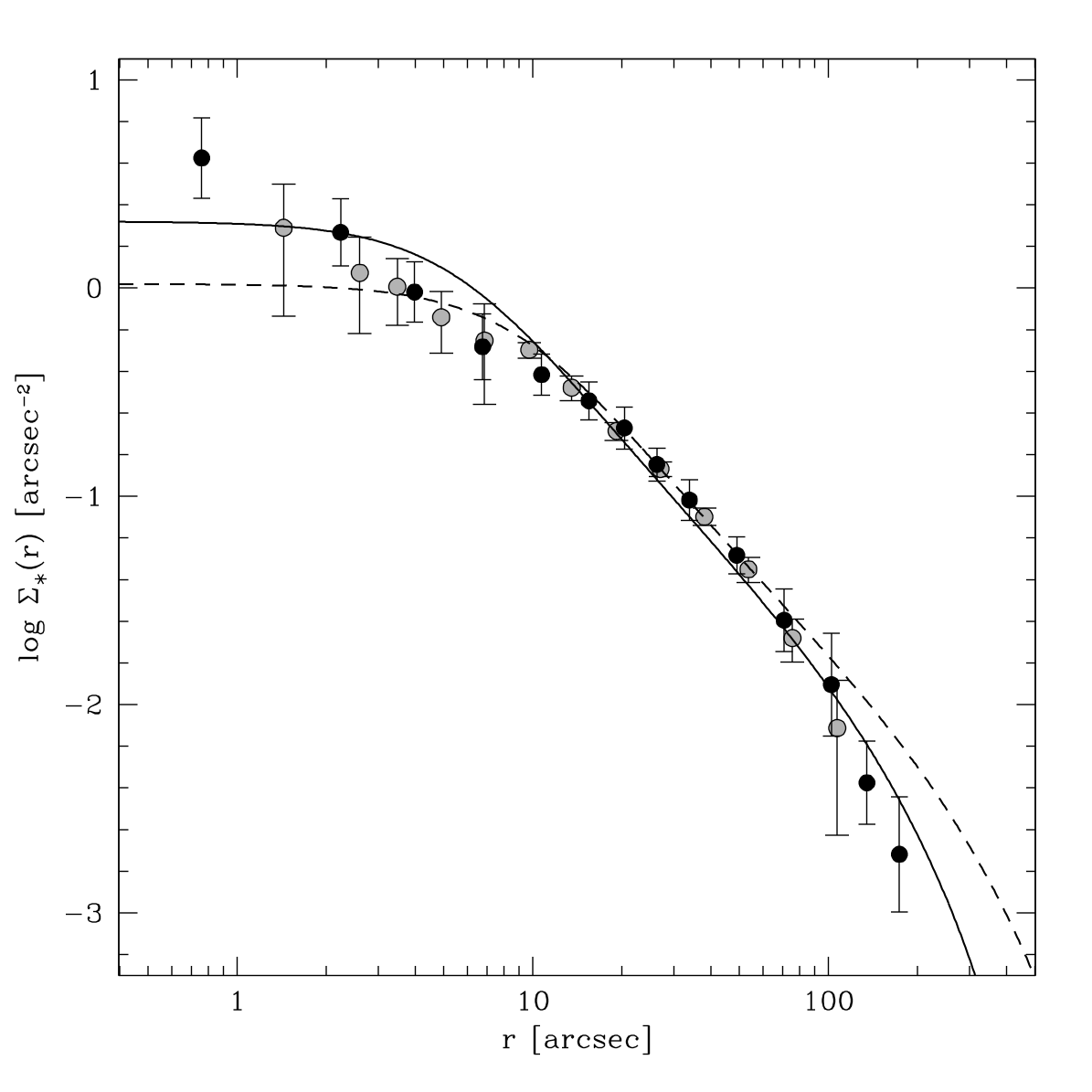}
\caption{Comparison between the background subtracted density profile
  of NGC 6256 determined in this work (black circles), and that
  obtained by \citet{cohen21}, vertically rescaled to match the former
  (grey circles). The solid and dashed lines are, respectively, the
  best-fit King models determined here and in \citet{cohen21}.}
\label{fig:cfr_cohen}
\end{figure}

The density profile has been obtained using only TO, sub-giant and
red-giant stars, which approximately have the same mass. Therefore,
the cluster's structural parameters have been derived by fitting the
observed profile with a single-mass \citet{king66} model, assuming
spherical symmetry and orbital isotropy.  Following \citet{raso20}, we
performed the fit using a Markov Chain Monte Carlo approach by means
of the {\rm emcee} package \citep{emcee13,emcee19}. We assumed uniform
priors on the parameters of the fit (i.e. the King concentration
parameter $c$, the core radius $r_c$, and the value of the central
density). Therefore, the posterior probability distribution functions
are proportional to the likelihood $\mathscr{L}=\exp{(-\chi^2/2)}$,
where the $\chi^2$ statistic is calculated between the measured
density values and those predicted by the whole family of adopted
models.  {As typically found for core collapsed GCs
  \citep[e.g.,][]{ferraro03_6752, zocchi16}, the resulting best-fit
  model (red dashed curve in the left-hand panel of Figure 3) clearly
  fails in properly reproducing the observations, and it is
  characterized by a value of the concentration parameter $c$ larger
  than 2.  Instead, a very good King model fitting is obtained if the
  three innermost points ($r < 5\arcsec$) are excluded from the
  analysis (red dashed curve in the right-hand panel of Figure 3). In
  turn, the evident central density cusp is well reproduced by a
  straight line, with a steep slope $\alpha_{\rm CUSP} = -0.89$ (red
  solid line in the figure).}  This is indeed the typical behavior of
a PCC cluster \citep[e.g.][]{vesperini10}.  The main structural
parameters obtained from best-fit King model are also labelled in the
figures, where $W_0$ is the model dimensionless potential, $r_{hm}$ is
the 3D half-mass radius, and $r_t$ is the truncation or tidal radius.
Figure \ref{fig:cfr_cohen} compares the density profile obtained in
this work (black circles) with that determined through a very similar
procedure by (\citealt{cohen21}, grey circles), vertically rescaled to
match our data using an average density difference value.  {The
  two profiles are in overall good agreement, the main difference
  being that we sample a larger radial extension, both toward the
  center and in the periphery, with the density that keeps increasing
  in the innermost bin, as expected in the presence of a power-law
  cusp. This difference is likely due to different adoptions of the
  magnitude selection and slightly different coordinates of the
  cluster center. Nevertheless, the structural parameters derived here
  and in Cohen et al. (2021) are in agreement within $\sim 1-2 \sigma$
  errors.}

\section{The BSS population and the $A^+$ parameter}
\label{sec:bss}
The shape of the density profile clearly indicates that NGC 6256 is in
a very advanced stage of its dynamical evolution, having already
experienced the CC phase. To independently and more quantitatively
investigate the dynamical status of this system, we analyzed the
properties of the BSS population, which are known to act as a
``dynamical clock'' \citet{ferraro12, ferraro18, ferraro19, ferraro20,
  lanzoni16}). However, NGC 6256 suffers from severe contamination
from Galactic interlopers \citep{cadelano20}, and a solid
characterisation of the BSS population first requires a proper
distinction between cluster's members and field objects.
Unfortunately, due to crowding, large extinction and distance, the
kinematic information provided by the Gaia DR3 can be used only for
the brightest portion of the red giant branch. Therefore, we took
advantage of the large temporal baseline between the two HST data-sets
to perform the PM analysis of the stars in common between the two
epochs. A similar, independent analysis was also performed by
\citet{cohen21}.

\subsection{Proper motion selection}
\label{sec:pm}
We adopted the approach described in \citet[][see also
  \citealp{bellini14, cadelano17, dalessandro18,
    raso20}]{dalessandro13}. The procedure consists in measuring the
instrumental position displacements of the stars detected in both
epochs, once a common distortion-free reference frame is defined. For
each data-set, we derived the mean instrumental positions ($x,y$) as
the $\sigma$-clipped mean of the positions of stars detected in at
least half the total number of images. In the WFC3 case, the resulting
($x,y$) positions have been corrected for geometric distortions by
applying the equations published in \citet{bellini11}. For the ACS
catalog, we adopted the ACS/WFC Distortion Correction Tables (IDCTAB)
provided in the dedicated page of the Space Telescope Science
Institute. The latter catalog was adopted as distortion-free reference
frame, due to its slightly larger FOV.  Then, we determined accurate
transformations between the first epoch WFC3 catalogue and the
reference frame. To this aim, we selected in both the catalogs a
sample of stars that can be considered as likely cluster members on
the basis of their CMD positions. We then applied a six-parameter
linear transformation\footnote{To do this, we used CataXcorr, a code
developed by P. Montegriffo at INAF—Osservatorio di Astrofisica e
Scienze dello Spazio di Bologna. This package is available at
\url{http://davide2.bo.astro. it/?paolo/Main/CataPack.html}, and has
been successfully used in a large number of papers by our group in the
past years.} to transform the positions of stars in the WFC3 frame to
the reference frame, treating each chip independently in order to
maximise the accuracy.  The derived transformations have been then
applied to all the stars in common between the two catalogues. The
relative PMs are finally determined by measuring the difference of the
mean ($x,y$) positions of each star in the two epochs, divided by
their temporal baseline and multiplied by the pixel scale of the
reference frame (50 mas pixel$^{-1}$). In such a way, the PMs along
both the right ascension ($\mu_{\alpha}\cos{\delta}$) and the
declination ($\mu_{\delta}$) are expressed into units of mas
yr$^{-1}$.

\begin{figure}[h] 
\centering
\includegraphics[scale=0.45]{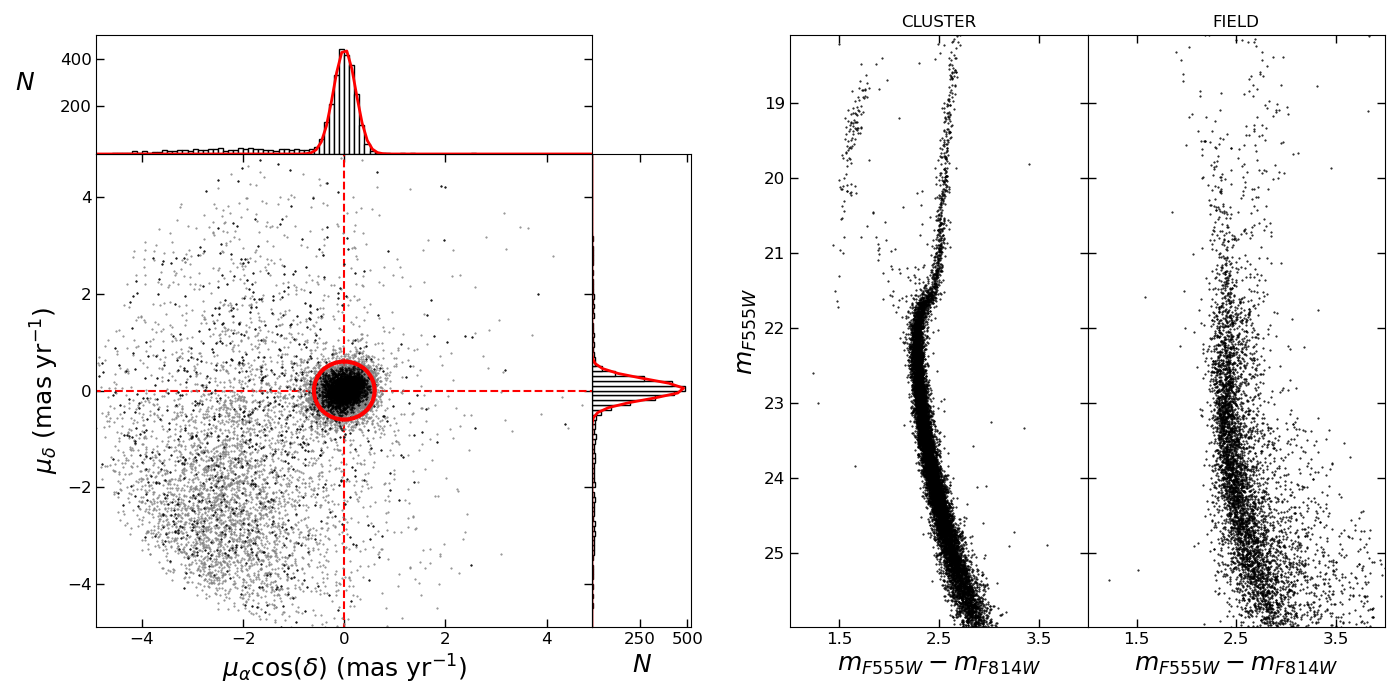}
\caption{{\it Left panel:} VPD of the stars in common between the two
  HST epochs, with the black dots being the sub-sample of stars with
  magnitude $19.0<m_{\rm F555W}<22.5$, and the gray dots corresponding
  to all the remaining stars. The upper and side plots show the
  histograms of the PM distribution of the selected sub-sample along
  the RA and Dec directions, with superimposed the best-fit Gaussian
  function. The red circle in the main panel is centered on the mean
  values of the best-fit Gaussian functions (i.e., 0 mas yr$^{-1}$)
  and has a radius of 0.6 mas yr$^{-1}$, equal to the combined
  $2\sigma$ PM dispersion of the sub-sample (see text). This circle
  encloses all the stars that have been selected as bona-fide cluster
  members.  {\it Right panels:} CMDs of the first epoch data-set for
  all the stars selected as cluster members (left) and for the stars
  selected as field interlopes (right).}
\label{fig:vpdcmd}
\end{figure}

The resulting Vector Point Diagram (VPD) is shown in the left panel of
Figure~\ref{fig:vpdcmd}. To maximize the efficiency in removing field
interlopers in the BSS region, we selected a sample of stars with
$19.0<m_{\rm F555W}<22.5$ and evaluated their PM distributions along
both the directions (see histograms in Figure~\ref{fig:vpdcmd}). These
distributions have been fitted with Gaussian functions centered on 0
and with standard deviation $\sigma\sim0.21$ mas yr$^{-1}$. We
selected as bona-fide cluster members those stars having a total PM
smaller than the combined $2\sigma$ dispersion (i.e., 0.6 mas
yr$^{-1}$). The resulting CMDs showing separately cluster's and field
members after the PM selection are shown in the two rightmost panels
of Figure~\ref{fig:vpdcmd}.

\begin{figure}[h] 
\centering
\includegraphics[scale=0.25]{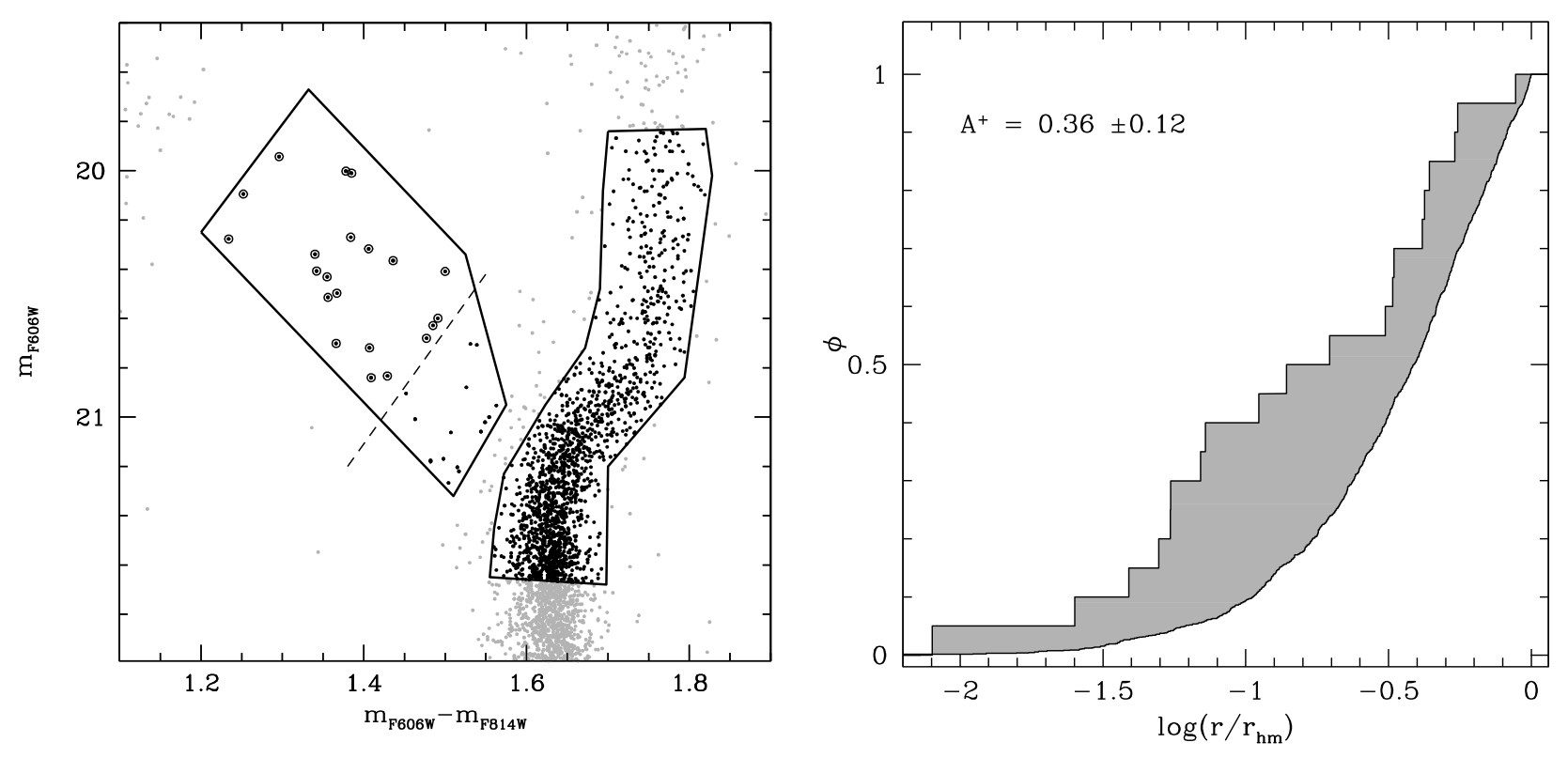}
\caption{{Left panel: PM cleaned CMD of NGC 6256 (grey dots) with
    the selected populations of 37 BSSs and 1585 reference stars
    highlighted in black. The solid lines draw the adopted selection
    boxes. The dashed line marks the threshold used to select the
    bright BSS sample (empty circles) for the computation of
    $A^+$. Right panel: Cumulative radial distributions of the 20
    bright BSSs (upper line) and the 1368 reference stars (lower line)
    included with $r_{hm}$, used to determine the $A^+$ parameter. The
    area of the region shaded in grey between the two cumulative
    radial distributions is the value of $A^+_{rh}$, which is also
    labelled in the panel.}}
\label{fig:cmdsel}
\end{figure}

The decontamination of the CMD through PM selection successfully
disentangles the cluster's population from the field one, thus opening
the possibility to study in detail the properties of the BSSs
population in the cluster's regions covered by the HST observations.
To this purpose, we used the ACS photometry that provides the deepest
and largest number of exposures, thus granting a significantly larger
signal to noise ratio in the CMD region occupied by BSSs, with respect
to what achievable with the WFC3 data. The samples of {37 BSSs and
  1585 reference stars} used in the following analysis have been
selected from the PM-cleaned CMD as shown in Figure
\ref{fig:cmdsel}. {To draw the BSS selection box, special care has
  been devoted to separate this population from MS-TO and SBG stars,
  and avoid the inclusion of MS-TO star blends. To this end, we built
  the color histograms of the measured stars in different bins of
  magnitudes $\sim 0.2$ wide, and we set the lower boundary of the BSS
  box at more than $4 \sigma$ from the SGB distribution. The resulting
  box is similar to, but not coincident with, that of
  \citet{leigh11}. Indeed, we adopted more conservative limits to
  avoid the inclusion of spurious objects, like (very bright)
  evolved-BSSs and, especially, photometric blends of MS-TO stars.}
The reference population is composed of MS-TO, sub-giant branch and
red giant branch stars in a magnitude range similar to that occupied
by the BSSs.

\subsection{A double sequence of BSSs}
\label{sec:double_seq}
\begin{figure}[h] 
\centering \includegraphics[scale=0.35]{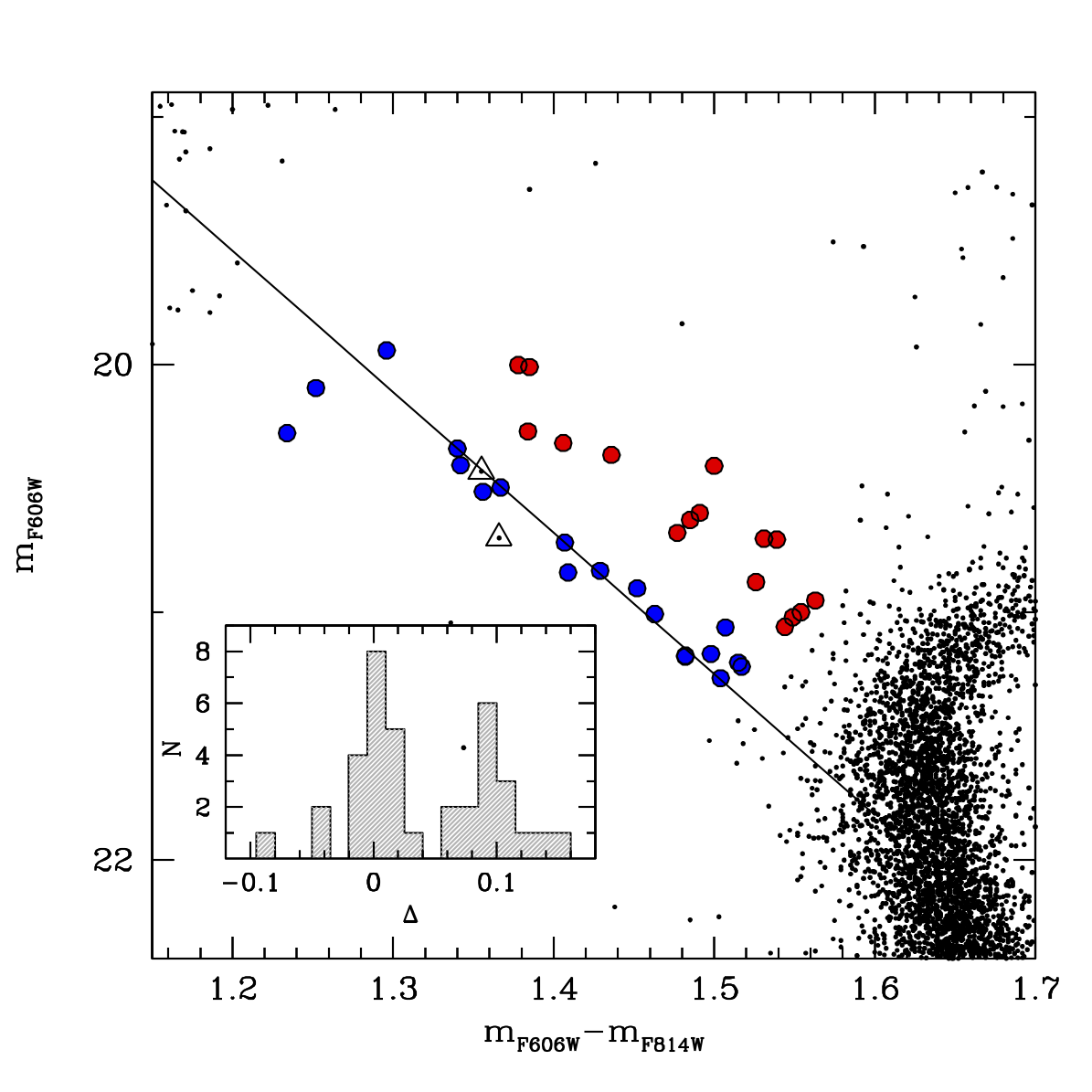}
\caption{{PM-selected CMD of NGC 6256 zoomed into the BSS region, with
  the BSSs aligned along the blue and the red sequence plotted as blue
  and red large circles, respectively.  The empty triangles mark
  candidate variable stars. The black line is the linear fit to the
  blue BSS sequence (with slope 4.7 and intercept 14.04). The
  distribution of the BSS distances from this line is plotted as a
  histogram in the inset panel.}}
\label{fig:bss_histo}
\end{figure}

Figure \ref{fig:bss_histo} shows the cluster CMD zoomed in the BSS
region.  It is clearly populated by a narrow sequence of 21 BSSs on
the bluer side, and by a more sparse population at redder colors,
composed of 16 objects. {It is worth mentioning that this result
  does not change if we apply different, less conservative PM
  selections. As a matter of fact, the double BSS sequence is clearly
  detected also in a CMD where no PM selection is applied at all (in
  that case, 6 more BSSs would be selected within the box in
  Figure~\ref{fig:cmdsel}, and they still aligne along the blue and
  red sequences).}  In order to quantitatively investigate the BSS
distribution in the CMD, we measured the distance of each star from
the best linear fit to the blue sequence ({black line} in the
figure). The histogram of the BSS distances from this line is shown in
the inset panel and it unambiguously reveals the presence of {two
  well-defined peaks separated by $\sim 0.1$ mag (i.e., almost one
  order of magnitude larger than the typical photometric color error
  in this magnitude range, which ranges between 0.014 and 0.016). Both
  the AIC and the BIC tests \citep[e.g.,][]{hastie01} confirm that
  the observed histogram is best reproduced by the sum of two Gaussian
  functions, with a probability larger than 99.5\% that the
  distribution is bimodal, instead of unimodal.}  This is reminiscent
of the behavior observed in other PCC clusters, where the narrow blue
sequence has been interpreted as the evidence of a collisional
population of BSSs (e.g., \citealt{ferraro09}; also see
Section~\ref{sec:intro}).

\begin{figure}[h] 
\centering \includegraphics[scale=0.35]{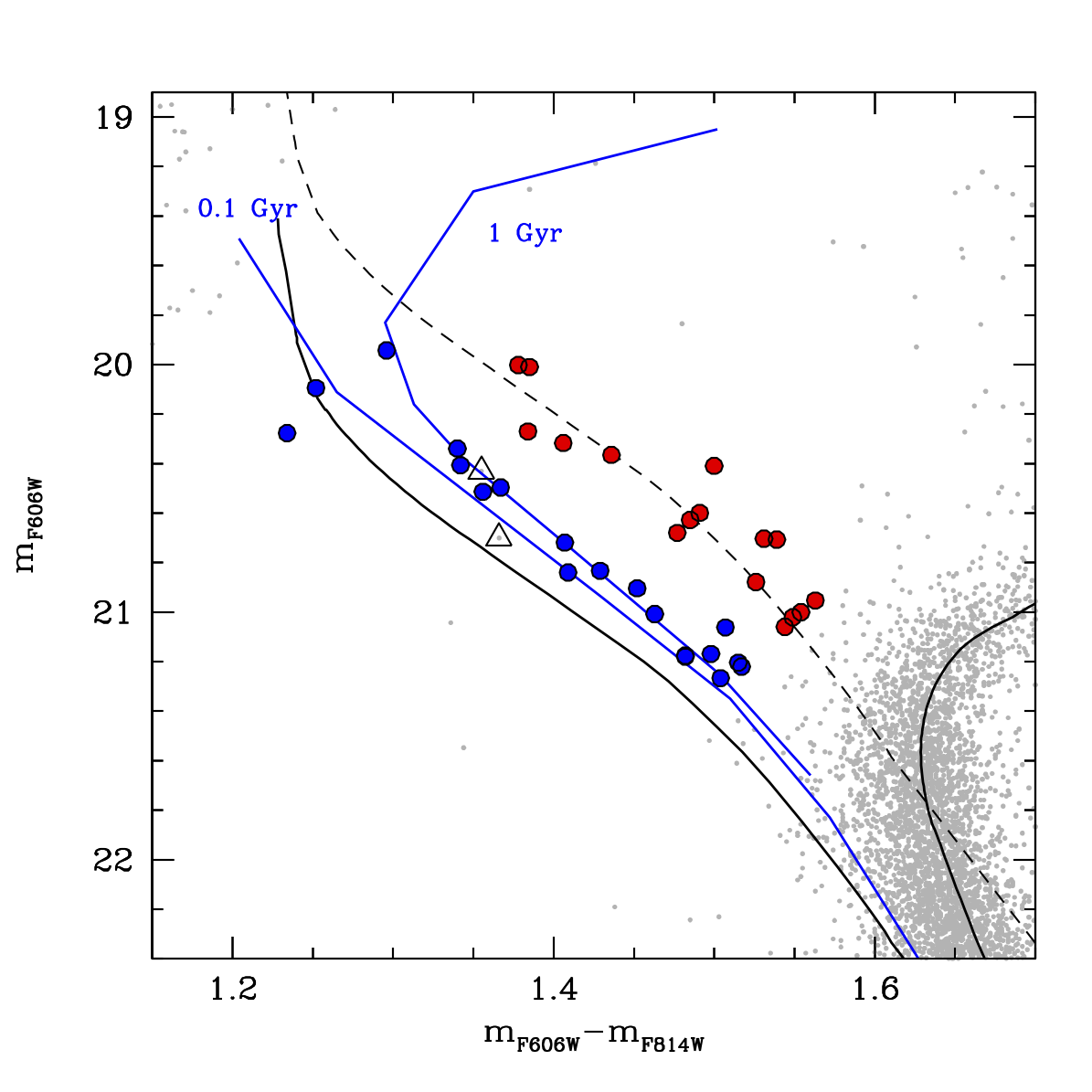}
\caption{{As in Figure \ref{fig:bss_histo}, but with models superposed.
  The two thick blue lines correspond to collisional isochrones
  \citep{sills09} of 0.1 and 1 Gyr (see labels) and they well repoduce
  the observed BSS blue sequence.  The black solid lines are standard
isochrones extracted from the \citet{dotter08} database and computed for the cluster
  metallicity ([Fe/H]$=-1.6$) at 1 and 13 Gyr (leftmost and rightmost
  curves, respectively). The black dashed lines corresponds to the 1
  Gyr isochrone shifted in magnitude by $=-0.75$.}}
\label{fig:coll_isoc}
\end{figure}

To verify whether this can be the case also for NGC 6256, we compared
the observations with a set of collisional models of intermediate
metallicity ($[Fe/H]=-1.3$) from the \citet{sills09} database.
Shortly, collisional models adopt the Yale Rotational Evolutionary
Code (YREC; \citealp{guenther92}) and have been used to simulate and
follow the evolution of a set of prototype BSSs formed by direct
collisions between two MS isolated stars. In particular, 16 cases
involving collisions of $0.4 M_\odot$, $0.6 M_\odot$ and $0.8 M_\odot$
stars are investigated. Collisional products are generated by using
the code ``Make Me A Star'' \citep{lombardi02} by assuming that the
collision occurs with a periastron separation of 0.5 times the sum of
the radii of the two stars.  The evolution of collision products from
the end of the collision to the MS phase is then traced following the
prescription by \citet{sills97}, and {and it is stopped when the
  energy generation due to hydrogen burning is larger than that due to
  gravitational contraction, which corresponds to the zero-age MS. The
  subsequent evolution on the MS, giant, horizontal and asympthotic
  giant branches is finally followed by using the Monash stellar
  evolution code for normal, low-mass stars \citep{karakas02}. Two
  collisional isochrones (at 0.1 and 1 Gyr) have been determined from
  the evolutionary tracks of three collision events involving stars
  with different masses ($0.6M_\odot+0.8M_\odot$,
  $0.6M_\odot+0.6M_\odot$ and $0.5M_\odot+0.6 M_\odot$), and they are
  shown as blue curves in Figure \ref{fig:coll_isoc}.} The matching
with the blue BSS sequence is impressive: the position of the vast
majority of the BSSs observed on the blue side of the CMD is very
nicely reproduced by the 1 Gyr old collisional isochrone.  NGC 6256 is
therefore the fourth PCC cluster where a sequence of collisional BSSs
is clearly identified. {The dashed line in the figure
  (corresponding approximately to the zero-age MS shifted by $-0.75$
  magnitudes) traces the locus occupied by (unresolved) equal-mass
  binary systems and it well corresponds to the CMD position of the
  red BSS sequence.}  According to the scenario depicted in Section
\ref{sec:intro}, BSSs belonging to the blue sequence are mostly formed
through collisions in a recent and short-lasting event (the cluster
CC), while those belonging to the red sequence are likely formed
through MT in binary systems.

The CMD region occupied by BSSs can be populated by variable stars
\citep[e.g.][]{rodriguez00, dieball07, beccari19}, with the optical
variability due either to pulsations (as in the case of SX Phoenicis),
or to binarity (e.g., cataclysmic variables and W-Uma).  To verify
whether some of the selected BSSs belongs to these categories, we
first inspected the Catalogue of Variable Stars in GCs by
\citet{clement01}. This lists a bright pulsating variable \citep[see
  also][]{matsunaga06} in the direction of NGC 6256, with a mean
K-band magnitude of 10.85, which is however saturated in our
exposures. We also verified that are no common stars among our BSSs
catalog and {the GAIA DR3 catalog}, which provides a flag for
stars displaying photometric variability. The lack of BSSs in common
with Gaia is not surprising and it is due to limited performances of
the satellite in the crowded and highly extincted environment of this
cluster. Finally, we carefully investigated the possible presence of
optical variability along the BSS sequence by analyzing the single
frame magnitudes obtained from our photometric analysis.  Two stars
show significant magnitude variations due to intrinsic variability
rather than photometric errors. They both lie along the blue sequence
(empty triangles in Figure~\ref{fig:coll_isoc}), where BSSs formed
though collisions (hence, not belonging to binary systems) are
expected.  However, to confirm the variability of these objects and
shed light on its nature (pulsation or binarity) a larger number of
observations able to properly sample the light curve is needed.  In
any case, {a very small} contamination of the blue (collisional)
sequence from variable binary stars has been already observed in other
clusters. For example, {one W-Uma variable has been identified
  along the blue BSS sequence in both} M30 \citep{ferraro09} and NGC
362 \citep{dalessandro13}.

{No statistically significant difference is found between the
  cumulative radial distributions of the blue and the red
  sequences. Conversely, the red BSS population was found to be more
  centrally segregated than the blue one in the other investigated
  GCs, with a large ($3 \sigma$) significance only in the case of NGC
  362 \citep{dalessandro13}, where the samples are the most
  numerous. This finding could therefore be due just to an effect of
  small statistics, or it could hide some clues about the formation of
  these objects. More similar investigations in a larger sample of GCs
  are needed to clarify this issue and properly address the physical
  origin of the possible difference in the radial segregation of the
  two BSS populations.}

\subsection{Measuring the dynamical age from the “dynamical clock" }
\label{sec:apiu}
To further investigate the dynamical status of NGC 6256, we applied
the so-called “dynamical clock": an empirical measurement of the
cluster dynamical evolution level based on the observational
properties of the BSS \citep{ferraro12,ferraro18}.  In particular,
following \citet{lanzoni16,ferraro18,ferraro20}, we measured the
segregation level of the BSS as determined by the value of the
$A^+_{rh}$ parameter, defined by \citet{alessandrini16} as the area
between the cumulative radial distributions of BSSs and reference
stars selected within one half-mass radius from the center. For a
proper comparison with previous results obtained for $\sim 1/3$ of the
entire Galactic GC population \citep{ferraro18}, we selected only the
brighter portion of the BSS population {(i.e., the stars above the
  dashed line shown in the left panel of Figure \ref{fig:cmdsel}).}
We adopted the half-mass radius derived from the best-fit with a King
model of the entire density profile ($r_{hm} = 73\arcsec$; left panel
of Fig. \ref{fig:densprof}). Although this value and that obtained
excluding the innermost $5\arcsec$ from the fit are in agreement
within the uncertainties, the former, in spite of a poorer fit to the
observations, better corresponds to the true distance from the center
that includes half of the total cluster mass (see right panel of
Fig. \ref{fig:densprof}).  {The right panel of
  Figure~\ref{fig:cmdsel} shows the cumulative radial distributions of
  the 20 BSSs and 1368 reference stars (upper and lower black lines,
  respectively) selected within the half-mass radius.} As expected for
dynamically evolved GCs, the BSS distribution is clearly more
centrally concentrated than the reference sample. {Indeed, the
  Kolmogorov-Smirnoff (K-S) and the Anderson-Darling (A-D) tests
  return a probability of 0.008 and 0.001, respectively, that the two
  samples are extracted from the same parent distribution.}

\begin{figure}[h] 
\centering
\includegraphics[scale=0.5]{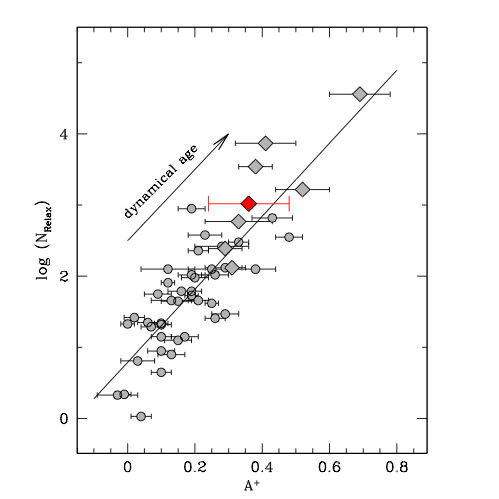}
\caption{{Relation between $A^+$ and $\log(N_{\rm relax})$ (solid
    line) obtained by \citet{ferraro18} for 48 Galactic GCs (grey
    symbols: circles and diamonds mark ``normal'' and PCC clusters,
    respectively). The position of NGC 6256 is marked with a red
    diamond.}}
\label{fig:apiu}
\end{figure}

The area between the two cumulative distributions shown in
Figure~\ref{fig:cmdsel} is shaded in grey and corresponds to
$A^+_{rh}=0.36\pm 0.12$.  Since the main source of uncertainty is the
small number statistics of the BSS population, the error has been
estimated through a jackknife bootstrapping technique \citep[][see
  also \citealp{dalessandro19}]{lupton93}, recalculating the area
between the cumulative distributions $N_{\rm BSS}$ times (where
$N_{\rm BSS}=20$ is the number of BSSs), excluding during each
iteration one star from the sample. The obtained value of $A^+_{rh}$
is among the largest determined so far (see \citealp{ferraro18}) for
Galactic GCs.  This is fully consistent with the classification of NGC
6256 as a PCC cluster. It is worth mentioning that, as shown in
Figure~\ref{fig:map}, the available HST data uniformly samples about
$\sim84\%$ of the cluster region within one half-mass radius, the
remaining $\sim16\%$ is outside the FOV beyond the edges of the ACS
camera. This could lead to a slightly underestimate of the $A^+_{rh}$
value. With this caveat in mind, we compared the position of NGC 6256
in the $A^+_{rh}$ versus $N_{\rm relax}$ diagram, together with the 48
Galactic GCs already investigated by \citet{ferraro18}.  As in
previous studies, $N_{\rm relax}$ is defined as the ratio between the
typical GC age (12 Gyr) and the central relaxation time ($t_{\rm rc}$)
of each system. This is usually estimated from equation (10) in
\citet{djorgovski93}, which is appropriate for stellar systems well
described by the King model family.  Although this is not the case for
NGC 6256 (see Section \ref{sec:profile}), we still adopted the same
approach assuming as concentration parameter and core radius the
values obtained from the fit to entire density profile (left panel in
Fig. \ref{fig:densprof}).  We adopted a central surface brightness
$\mu_0=18.36$, directly measured from the F555W images within
$1\arcsec$ from the center. This value has been corrected for the
effect of extinction using an average color excess of $E(B-V)=1.34$,
estimated from the reddening map of \citet{cadelano20} within the same
sky area.  We also adopted a distance $d=6.8$ kpc \citep{cadelano20},
an absolute magnitude $M_V=-7.15$ \citep[][2010 edition]{harris96}, an
average stellar mass of $0.3 M_\odot$ and a mass-to-light ratio of 2
{\bf(which are typical values for a $\sim13$ Gyr old GC; e.g.,
  \citealp{maraston98})}. The resulting value of the central
relaxation time is $\log(t_{\rm rc}) = 7.06$ (expressed in years),
indicating that the cluster experienced more than one thousand current
relaxation times since its formation, corresponding to $\log(N_{\rm
  relax})=3.02$. The position of NGC 6256 in the $A^+_{rh}$ versus
$N_{\rm relax}$ diagram is highlighted {as a red diamond} in
Figure \ref{fig:apiu}. Its position is in very good agreement with the
trend drawn by the Galactic GC population and confirms that NGC 6256
is in a very advanced stage of its dynamical evolution. {Indeed,
  it lies in the same region of the diagram occupied by the other PCC
  clusters of the sample, and its inclusion negligibly modifies the
  coefficients of the best-fit line.}  All this further demonstrates
that $A^+_{rh}$ is a powerful indicator of the degree of dynamical
evolution of GCs.

\section{Summary and Conclusions}
\label{sec:conclusions}
{In this paper we have performed a detailed photometric
  investigation of the bulge GC NGC 6256, also taking advantage of HST
  multi-epoch data to measure relative PMs in the inner region of the
  system for the accurate decontamination of the CMD from field
  stars. The analysis demonstrated that this GC is in a very advanced
  stage of dynamical evolution. In fact, the observed stellar density
  profile cannot be properly reproduced by a King model, and shows a
  steep power-law cusp with a slope of $-0.89$ in the innermost $\sim
  5\arcsec$.  This is the typical behavior expected for a cluster that
  already experienced CC \citep[see e.g.][]{bhat22}. A very advanced
  dynamical stage is also suggested by the value of the $A^+$
  parameter \citep{alessandrini16}, measuring the level of BSS central
  segregation, with respect to lighter (MS-TO and giant) stars. As
  shown in Fig. \ref{fig:apiu}, NGC 6256 well follows the same trend
  between $A^+$ and $N_{\rm relax}$ traced by $\sim 1/3$ of the entire
  Galactic GC population \citep[from][]{ferraro18}, and it locates in
  region of the diagram occupied by PCC clusters (diamonds).

  The analysis also showed that the BSS population of NGC 6256 draws a
  well-defined narrow blue sequence (corresponding to what expected
  for a sub-population of BSSs with different masses generated over a
  relatively short time scale), and a more sparse red sequence.  This
  is similar to what observed in other PCC clusters (M30, NGC 362,
  M15), where it has been interpreted \citep[e.g.][]{ferraro09} as the
  manifestation of the two main BSS formation channels: the blue
  sequence should be originated mainly by collisions over a short
  time-scale, while the red sequence derives from a continuous
  formation process \citep[see also][]{portegies19}, as expected in
  the case of MT-generated BSSs. Indeed, the comparison with
  collisional models from \citet{sills09} computed for a metallicity
  appropriate for NGC 6256 provides a very good match to the
  observations: the narrow blue sequence is well reproduced by a
  collisional isochrone of 1 Gyr, thus suggesting that, approximately
  1 Gyr ago, a short-lasting event occurred and promoted the formation
  of a population of collisional BSSs, which currently represents
  $\sim60\%$ of the entire BSS cluster content.  The fact that NGC
  6256 is a PCC system naturally yields to the conclusion that the
  blue sequence was originated by the increased collision rate during
  the CC phase.  Based on analogous considerations, \citet{ferraro18}
  and \citet{dalessandro13} concluded that M30 experienced CC
  approximately 2 Gyr ago, and NGC 362 reached this phase of dynamical
  evolution more recently, about 0.2 Gyr ago. In the case of M15, the
  blue sequence is further separated in two different branches: the
  main one extends up to 2.5 mag above the MS-TO and is nicely
  reproduced by a $\sim2.2$ Gyr collisional isochrone, while the
  second is $\sim 1$ mag less extended in the CMD and reproduced by a
  $\sim 5.5$ Gyr collisional isochrone \citep{beccari19}. The authors
  suggest that such a complex feature could be the result of two
  distinct events of high collisional activity: the first one is the
  cluster CC, occurred about 5.5 Gyr ago, and the second and more
  recent one corresponds to a core oscillation during the PCC
  evolution (see Figure 9 in \citealp{beccari19}).  Intriguingly, the
  CMD positions of the two brightest BSSs along the blue sequence of
  NGC 6256 are consistent with a 0.1 Gyr collisional isochrone, thus
  possibly indicating the epoch of the last relevant re-collapse
  during the PCC gravothermal oscillation phase, similarly to the case
  of M15.

Of course, the photometric separation of the two kinds of BSSs in the
CMD is not completely stringent. In fact, while MT-BSSs are unable to
produce blue sequences as narrow and well-defined as observed, recent
MT models \citep{jiang17} demonstrate that some of them can
``contaminate'' the blue region of the CMD. This is indeed consistent
with the presence of W-Uma variables identified along the blue
sequence of M30 \citep{ferraro09} and NGC 362 \citep{dalessandro13},
and possibly also in NGC 6256, where hints of photometric variability
have been found for two blue BSSs (see Figure \ref{fig:coll_isoc}).
On the other hand, also collisional BSSs that were born along the blue
sequence cross the red sequence region during their natural
evolution. However, the possibility of observing a collisional BSS
during its evolution through the red portion of the CMD is low (for
instance, a $1.1 M_\odot$ collisional BSS originated by a
$0.5M_\odot+0.6 M_\odot$ collision spends approximativel 2 Gyr along
the blue sequence, and only a few $10^8$ yr crossing the red strip).

The analysis discussed in this paper for NGC 6256, together with
previous results for other PCC and ``normal'' GCs
\citep[e.g.][]{ferraro09, ferraro18, dalessandro13, beccari19},
provide excellent examples of the potential prediction power of the
BSS distribution in tracing the dynamical history of stellar
systems. On the one hand, the segregation level of BSSs is a very
powerful empirical way for the classification of GCs in terms of their
level of internal dynamical evolution.  On the other hand, the
morphology of the BSS sequence in the CMD can be used to trace the
time-scale of dynamical evolution during and after CC.}

\acknowledgments {We thanks the anonymous referee for all the
  comments and suggestions that improved the manuscript quality.} We
warmly thank Alison Sills for providing us the collision isochrones
used in this work.  This work is part of the project {\it Cosmic-Lab}
at the Physics and Astronomy Department "A. Righi" of the Bologna
University (http://www.cosmic-lab.eu/ Cosmic-Lab/Home.html). The
research was funded by the MIUR throughout the PRIN-2017 grant awarded
to the project Light-on-Dark (PI:Ferraro) through contract
PRIN-2017K7REXT.

\bibliography{biblio_bss}{}
\bibliographystyle{aasjournal}


\end{document}